\documentclass[%
 reprint,
 amsmath,amssymb,
 aps,
]{revtex4-1}

\usepackage{graphicx}
\usepackage{dcolumn}
\usepackage{bm}

\begin{document}

\preprint{APS/123-QED}

\title{Transmission of charge and spin in a topological-insulator-based magnetic structure}

\author{P. R. Rzeszutko}
\affiliation{%
	Department of Physics and Medical Engineering,
	Rzesz\'ow University of Technology, al.~Powsta\'nc\'ow Warszawy 6, 35-959 Rzesz\'ow, Poland
}%
\author{S. Kudla}%
 \email{s.kudla@prz.edu.pl}
\affiliation{%
Department of Physics and Medical Engineering,
Rzesz\'ow University of Technology, al.~Powsta\'nc\'ow Warszawy 6, 35-959 Rzesz\'ow, Poland
}%
\author{V. K. Dugaev}%
\email{vdugaev@prz.edu.pl}
\affiliation{%
Department of Physics and Medical Engineering,
Rzesz\'ow University of Technology, al.~Powsta\'nc\'ow Warszawy 6, 35-959 Rzesz\'ow, Poland
}%

\date{\today}

\begin{abstract}
{%
	We discuss the effect of a magnetic thin-film ribbon at the surface of a topological insulator on the charge and spin transport due to surface electrons.
}{%
	If the magnetization in the magnetic ribbon is perpendicular to the surface of a topological insulator, it leads to
	a gap in the energy spectrum of surface electrons. As a result, the ribbon is a barrier for electrons,	which leads to electrical resistance.
}{%
		We have calculated conductance of such a structure. The conductance
	reveal some oscillations with the length of the magnetized region due to the interference of 
	transmitted and reflected waves.
	We have also calculated the Seebeck coefficient when electron flux is due to a temperature gradient.
}
\begin{description}
\item[PACS numbers]
72.25.Dc, 72.25.Mk
\end{description}
\end{abstract}

\maketitle


\section{\label{sec:level1}Introduction}
Topological insulators (TIs) are systems which are insulators in the bulk, but their conducting properties are associated with surface
(in 3-dimensional case)    electronic states emerging in the gap~\cite{hasan10,qi11}. The unique electric and magnetic properties of TIs
are related mainly to robustness of the energy spectrum of surface electronic states to any perturbation which does not break the
time inversion symmetry~\cite{fu06,fu07}. This means that nonmagnetic impurities or defects do not open energy gap in
the Dirac spectrum of the surface electrons. On the other hand, the effect of magnetic field
or a nonzero magnetization at the surface can be crucial as they break the symmetry protection
mechanism of the topological insulator~\cite{liu09,zitko10,abanin11,cheianov12,rosenberg12,biswas10,zhang12}.
In particular, the magnetization oriented perpendicularly to the surface of a TI
opens an energy gap~\cite{liu09,zhang12,henk12,chen10,wray11}, whereas the gap is absent in the case of in-plane magnetization.
This behavior gives a chance to control the parameters of electron energy spectrum by variation of the magnetization
orientation, that can be done, for instance, by applying a weak external magnetic field.

Another important property of surface electrons described by the Dirac Hamiltonian is the total
transparency of any potential barrier for electrons incident on the barrier~\cite{katsnelson}. This phenomenon is known as the Klein effect.
However, formation of an energy gap
in the Dirac spectrum significantly changes the situation. This is because electrons can propagate only
{\it via} continuous energy spectrum, and are reflected from the gap.

In this paper we discuss the possibility of using a structure with a local magnetization region on top of a TI to control
transport of surface electrons. It is important that electron transport in a TI means
transport of charge and spin since the electron states are spin-polarized along the direction of electron motion.

It should be noted that structures of this type attracted a lot of attention recently 
\cite{kong11,yokoyama11,fan14,mellnik14,semenov14,chang15,mahfouzi16}  due to the possibility of an
effective control of magnetization in the magnetic layer by the electric current in TI. 
It is related to the current-generated spin
torque reversing the magnetic moments in the layer. Especially important is the spin-orbit component of the torque
\cite{kurebayashi14} which is due
to the current-induced spin polarization of electrons under spin-orbit interaction in the TI.

The other interesting types of a hybrid structure, which includes superconducting and magnetic layers
on top of the topological insulator, have been also considered \cite{tanaka09,dolcini15}.
In this case, one appear the Majorana excitations, which can be manipulated by the magnetization direction
\cite{tanaka09}, or a possibility of controlling the phase shift in the Josephson current by the in-plane
magnetic field \cite{dolcini15}.  

In this paper we mostly concentrate on charge and spin transmission through the magnetized region assuming that the
electric current is weak enough, so that the magnetic moments in the ribbon can be viewed as frozen. We consider
the model with perpendicular magnetization.  In this case the magnetization opens a gap in the electronic spectrum
of TI, which creates an effective barrier for the motion of spin-polarized elecrons in TI. Thus, by controlling
the magnetization orientation one can change the resistance of the structure \cite{kong11}.

\section{\label{sec:level1}Model}

We consider a system consisting of a three-dimensional (3D) topological insulator and a narrow thin-film magnetic ribbon on top of its surface.
Magnetization in the ribbon is assumed to be perpendicular to the surface of TI, as shown schematically
in Fig.~1. Hamiltonian which describes electrons at the surface of TI can be written in the following form:
\begin{eqnarray}
\label{1}
\hat{H}=-iv\left( \sigma _x \partial _x+\sigma _y\partial _y\right) +m(x)\, \sigma _z ,
\end{eqnarray}
where $\sigma _i$ are the spin Pauli matrices, while $m(x)=gM_z(x)$ leads to the energy gap
that is determined by the $z$-component of magnetization $M_z(x)$. The parameters $v/\hbar $ and $g$ are
the velocity of surface electrons in TI and the coupling constant, respectively.

\begin{figure}
	\hspace*{-0.2cm}
	\includegraphics[width=8.5cm]{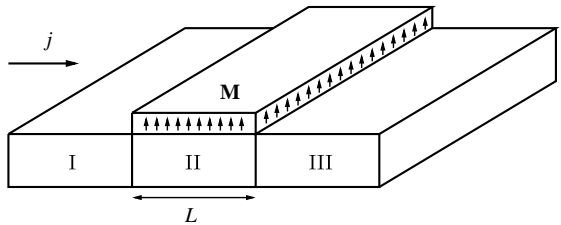}
	\caption{Schematic of a topological insulator with a magnetic thin-film ribbon on its surface.
		Magnetization ${\bf M}$ in the ribbon is perpendicular to the surface of TI.}
\end{figure}

Without any magnetic film at the surface, $m(x)=0$, the Hamiltonian (1) describes massless two-dimensional (2D) Dirac electrons
at the surface of TI. The energy spectrum of such electrons has two branches,
$\varepsilon _{1,2}(k)=\pm vk$. The main properties of 3D TIs are related to the fact that the low-energy
electron excitations are described by massless Dirac electrons, and the spectrum is
protected by the time-reversal symmetry of the system, so that any perturbation which does not
break this symmetry can not open an energy gap in the Dirac spectrum \cite{hasan10,qi11}.
As we can see from Eq.~(1), Dirac electrons
are spin-polarized along the wavevector ${\bf k}$, so the electric charge transmitted by Dirac electron
is accompanied by the transmission of spin oriented in the direction of electron motion. One can check
that any potential barrier can not stop the motion of massless Dirac electrons (Klein paradox \cite{katsnelson}).

The magnetization perpendicular to the TI surface can open the energy gap since the magnetization
breaks the time-reversal  invariance. As a result, electrons in TI can be effectively scattered and reflected
from the barrier. We assume that the magnetic region
of length $L$ is along the $x$-axis as shown in Fig.~1. Correspondingly, we can assume
$m(x)$ in the following form:
\begin{eqnarray}
\label{2}
m(x)=\left\{ \begin{array}{cc} m_0, & 0\le x\le L, \\ 0, & x<0\; {\rm or}\; x>L . \end{array}  \right.
\end{eqnarray}
The Schr\"odinger equation, $(\hat{H}-\varepsilon )\, \psi ({\bf r})=0$, for the spinor components $\varphi$,$\chi$ of
the wavefunction $\psi ({\bf r})=e^{ik_yy}\psi _{k_y}(x)$ can be presented as
\begin{eqnarray}
\label{3}
&&\left( m-\varepsilon \right) \varphi _{k_y}+v\left( -i\partial _x-ik_y\right) \chi _{k_y}=0,
\nonumber \\
&&v\left( -i\partial _x+ik_y\right) \varphi _{k_y}-\left( m+\varepsilon \right) \chi _{k_y}=0 .
\end{eqnarray}
Since we assumed the gap in the form given by Eq.~(2), these equations can be solved separately in the regions I, II and III
(see Fig.~1).

\section{\label{sec:level1}One-dimensional motion}

Let us consider first the motion of electrons along the axis $x$.
This corresponds to $k_y=0$ in equations (3).
By solving these equations in the regions I ($x<0$), II ($0<x<L$) and III ($x>L$) we find (similar formulae
are presented by Yokoyama \cite{yokoyama11} for a different choice of the Hamiltonian describing Bi$_2$Se$_3$ TI)
\begin{eqnarray}
\label{4}
\psi _{k_y=0}(x)=\frac{e^{ik_xx}}{\sqrt{2}}\left( \begin{array}{c} 1 \\ 1 \end{array}\right)
+\frac{re^{-ik_xx}}{\sqrt{2}}\left( \begin{array}{c} 1 \\ -1 \end{array}\right) ,\; x<0,
\end{eqnarray}
\begin{eqnarray}
\label{5}
\psi _{k_y=0}(x)=\frac{Ae^{ik'_xx}}{\sqrt{2\varepsilon (\varepsilon +m_0)}}
\left( \begin{array}{c} \varepsilon +m_0 \\ \sqrt{\varepsilon ^2-m_0^2} \end{array}\right) \hskip1.3cm
\nonumber \\
+\frac{Be^{-ik'_xx}}{\sqrt{2\varepsilon (\varepsilon +m_0)}}
\left( \begin{array}{c} \varepsilon +m_0 \\ -\sqrt{\varepsilon ^2-m_0^2} \end{array}\right) ,\; 0<x<L,
\end{eqnarray}
\begin{eqnarray}
\label{6}
\psi _{k_y=0}(x)=\frac{te^{ik_xx}}{\sqrt{2}}\left( \begin{array}{c} 1 \\ 1 \end{array}\right) ,\; x>L,\hskip1.8cm
\end{eqnarray}
where $r$ and $t$ are the coefficients  of reflection and transmission, $A$ and $B$ are constants,
$k_x=\varepsilon /v$, $k'_x=\sqrt{\varepsilon ^2-m_0^2}/v$, and we assumed $\varepsilon >m_0$.

Equation (4) represents the incoming and reflected waves with spin polarization along the
direction of the electron motion. Due to external magnetization ${\bf M}$ in the region II, the waves propagating in opposite directions
in this region (5) are not spin polarized in the direction of motion. If the energy of an electron is slightly above the
gap, $(\varepsilon -m_0)\ll m_0$, the effect of magnetization field is so strong that the electron
is spin polarized along axis $z$, i.e. perpendicular to the direction of electron motion. The transmitted (outgoing)
electron (6) is spin polarized along the axis $x$.

Matching the solutions (4)-(6) at the interfaces $x=0$ and $x=L$, i.e.,
$\psi _{k_y=0}(-\delta )=\psi _{k_y=0}(+\delta )$ and $\psi _{k_y=0}(L-\delta )=\psi _{k_y=0}(L+\delta )$,
we obtain four equations for the spinor components, and finally we can find the values of all constants $r,t,A,B$
entering Eqs.~(4)-(6). In particular, we find the transmission coefficient for an electron with energy $\varepsilon >m_0$
\begin{eqnarray}
\label{7}
t_{k_y=0}=4 e^{i(k'_x-k_x)L} \sqrt{\varepsilon ^2-m_0^2}
\left[ \left( \sqrt{\varepsilon +m_0}+\sqrt{\varepsilon -m_0}\right) ^2
\right. \nonumber \\ \left.
-e^{2ik'_xL}\left( \sqrt{\varepsilon +m_0}-\sqrt{\varepsilon -m_0}\right) ^2\right] ^{-1} .\hskip0.5cm
\end{eqnarray}

From this equation follows that the probability of transmission through the "magnetization
barrier", $T_{k_y=0}=|t_{k_y=0}|^2$, can go to zero if the electron energy is close to the value of the gap,
$(\varepsilon -m_0)\ll m_0$ and $L\to \infty $.

\begin{figure}
	\vspace*{-0.7cm}
	\hspace*{-1cm}
	\includegraphics[width=9cm]{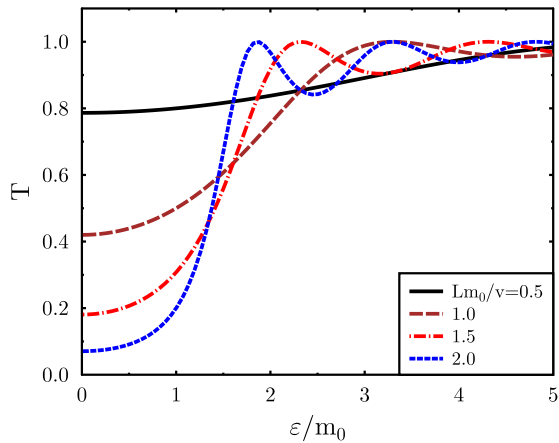}
	\caption{Transmission probability $T_{k_y=0}$ as a function of energy for different length of magnetization region.
		Electron is moving along axis $x$.}
\end{figure}

Indeed, in the limit of $(\varepsilon -m_0)\ll m_0$ from Eq.~(7) follows
\begin{eqnarray}
\label{8}
T_{k_y=0}\simeq \frac{4\, (\varepsilon -m_0)}
{m_0 (1-\cos 2k'_xL)+2(\varepsilon -m_0)(1+\cos 2k'_xL)}\, ,\hskip0.2cm
\end{eqnarray}
which means that the transmission probabilty $T_{k_y=0}$ is oscillating with the length $L$. 
These oscillations are due to the interference of transmitted and reflected waves in the magnetized region. If $k'_xL$ is
not very close to $n\pi $ ($n$ is integer), then we obtain $T_{k_y=0}=0$ but at the points $k'_xL=n\pi $
we get $T_{k_y=0}=1$. Since $k'_x=\sqrt{\varepsilon ^2-m_0^2}/v$, for
$(\varepsilon -m_0)\ll m_0$ and small enough $L\ll v/\sqrt{2m_0(\varepsilon -m_0)}$
we also obtain full transparency of the magnetic barrier, $T_{k_y=0}\simeq 1$.
It should be also mentioned that when $T_{k_y=0}=0$, the constants $A$ and $B$ in Eq.~(5) do not vanish,
which indicates that an electron coming from the region I can partially penetrate into the region II, but it does not
go through the outer interface $x=L$.
In the case of large electron energy, $\varepsilon \gg m_0$, we find from Eq.~(7) that $T_{k_y=0}\simeq 1$,
like in the case of TI without any magnetic barrier.

The electrons with energies $\varepsilon <m_0$ can penetrate through the gap by tunneling.
In this case, the wavefunction in region II has the following form
\begin{eqnarray}
\label{9}
\psi _{k_y=0}(x)=\frac{Ae^{\kappa x}}{\sqrt{2m_0 (\varepsilon +m_0)}}
\left( \begin{array}{c} \varepsilon +m_0 \\ -i\sqrt{m_0^2-\varepsilon ^2} \end{array}\right)
\nonumber \\
+\frac{Be^{-\kappa x}}{\sqrt{2m_0 (\varepsilon +m_0)}}
\left( \begin{array}{c} \varepsilon +m_0 \\ i\sqrt{m_0^2-\varepsilon ^2} \end{array}\right) ,\; 0<x<L,
\end{eqnarray}
where $\kappa =\sqrt{m_0^2-\varepsilon ^2}/v$. Hence, calculating the transmission coefficient like
before we find for $0<\varepsilon <m_0$
\begin{eqnarray}
\label{10}
t_{k_y=0}=-4i e^{(\kappa -ik_x)L} \sqrt{m_0^2-\varepsilon ^2}
\nonumber \\ \times
\left[ \left( \sqrt{m_0+\varepsilon }-i\sqrt{m_0-\varepsilon }\right) ^2
\right. \nonumber \\ \left.
-e^{2\kappa L}\left( \sqrt{m_0+\varepsilon }+i\sqrt{m_0-\varepsilon }\right) ^2\right] ^{-1} .
\end{eqnarray}
Using this expression (10) we can find that for $\varepsilon \ll m_0$ the transmission probability
has a simple form
\begin{eqnarray}
\label{11}
T_{k_y=0}=\frac2{1+\cosh (2\kappa L)}\; ,
\end{eqnarray}
and in the limit of $\kappa L\to 0$ we get $T_{k_y=0}=1$.

\begin{figure}
	\vspace*{-0.7cm}
	\hspace*{-1cm}
	\includegraphics[width=9cm]{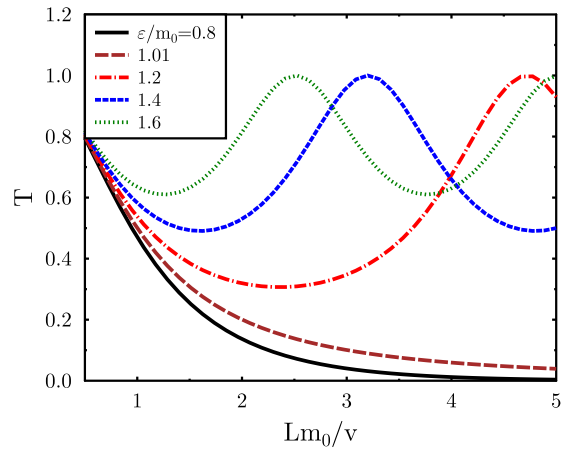}
	\caption{Transmission probability $T_{k_y=0}$ as a function of length $L$ of the magnetized region
		for different values of electron energy. Electrons are moving along the axis $x$.}
\end{figure}

The dependence of transmission probabilty $T_{k_y=0}$ on the electron energy $\varepsilon $ and on the
length $L$ are presented in Figs.~2 and 3, respectively. Here we used the dimensionless energy
$\varepsilon /m_0$ and length $Lm_0/v$. If we take $m_0=100$~meV and $v=10^{-8}$~eV$\cdot $cm,
we obtain the value of length unit $v/m_0=1$~nm.

\section{\label{sec:level1}Two-dimensional motion of electrons}

Let us consider first overbarrier transmission of electrons.
Taking into account a nonzero value of $k_y$ for electrons incoming at a certain angle
to the surface, we write the solution of Eqs.~(3) for the wavefunction in the regions I, II, and III as
(we omit the common multiplier $e^{ik_yy}$)
\begin{eqnarray}
\label{12}
\psi _{\bf k}(x)=\frac{e^{ik_xx}}{\sqrt{2}}\left( \begin{array}{c} 1 \\ k_+/k \end{array}\right)
+\frac{re^{-ik_xx}}{\sqrt{2}}\left( \begin{array}{c} 1 \\ -k_-/k \end{array}\right) ,\; x<0,\hskip0.5cm
\end{eqnarray}
\begin{eqnarray}
\label{13}
\psi _{\bf k}(x)=\frac{Ae^{ik'_xx}}{\sqrt{2\varepsilon (\varepsilon +m_0)}}
\left( \begin{array}{c} \varepsilon +m_0 \\ vk'_+ \end{array}\right) \hskip2cm
\nonumber \\
+\frac{Be^{-ik'_xx}}{\sqrt{2\varepsilon (\varepsilon +m_0)}}
\left( \begin{array}{c} \varepsilon +m_0 \\ -vk'_- \end{array}\right) ,\; 0<x<L,
\end{eqnarray}
\begin{eqnarray}
\label{14}
\psi _{\bf k}(x)=\frac{t\, e^{ik_xx}}{\sqrt{2}}\left( \begin{array}{c} 1 \\ k_+/k \end{array}\right) ,\; x>L,\hskip1.8cm
\end{eqnarray}
where we denote $k_\pm =k_x\pm ik_y$, $k'_\pm =k'_x\pm ik_y$, and $k'_x=(\varepsilon ^2-m_0^2-v^2k_y^2)^{1/2}/v$.
Here we take into account that the $k_y$ component of vector ${\bf k}$ is conserved when the electron is transmitted
through the interface. This is due to the translational invariance along $y$. Note that here we use vector ${\bf k}$
being the wavevector of incoming wave as a quantum number to label scattering states, which are the
eigenstates of Schr\"odinger equation with Hamiltonian (1).

After matching Eqs.~(12) to (14) at $x=0$ and $x=L$ one can obtain the following
expression for transmission coefficient depending on the direction of vector ${\bf k}$ for $\varepsilon >(m_0^2+v^2k_y^2)^{1/2}$
\begin{eqnarray}
\label{15}
t({\bf k})=2vk'_x\, e^{i(k'_x-k_x)L}(1+k/k_+)\, (\varepsilon +m_0) \hskip0.8cm
\nonumber \\ \times
\Big[ \Big( \varepsilon +m_0+\frac{\varepsilon k'_+}{k_-}\Big)
\Big( \varepsilon +m_0+\frac{\varepsilon k'_-}{k_+}\Big) -e^{2ik'_xL}
\nonumber \\ \times
\Big( \varepsilon +m_0-\frac{\varepsilon k'_+}{k_+}\Big)
\Big( \varepsilon +m_0-\frac{\varepsilon k'_-}{k_-}\Big) \Big] ^{-1}.
\end{eqnarray}
One can check that electrons incident on the interface at a nonzero angle $\phi =\tan ^{-1}(k_y/k_x)$ are not
transferred so easily as those with $\phi =0$.

One can also consider the subbarrier transmission of electrons.
In this case the electrons are tunneled through the gap, and instead of Eq.~(13) we get
\begin{eqnarray}
\label{16}
\psi _{\kappa ,k_y}(x)=\frac1{\sqrt{2[m_0 (\varepsilon +m_0)+v^2k_y(k_y-\kappa )]}}\hskip1cm
\nonumber \\ \times
\left[ Ae^{\kappa x}\left( \begin{array}{c} \varepsilon +m_0 \\ -iv(\kappa -k_y) \end{array}\right)
+Be^{-\kappa x}\left( \begin{array}{c} \varepsilon +m_0 \\ iv(\kappa +k_y) \end{array}\right) \right] ,
\nonumber \\
0<x<L, \hskip0.5cm
\end{eqnarray}
whereas for $x<0$ and $x>L$ we can use again Eqs.~(12) and (14).
Here we denote $\kappa=(m_0^2+v^2k_y^2-\varepsilon ^2)^{1/2}/v$.

Then for the transmission coefficient at $0<\varepsilon <(m_0^2+v^2k_y^2)^{1/2}$ we obtain
\begin{eqnarray}
\label{17}
t({\bf k})= \Big( 1+\frac{k_+}{k_-}\Big) \, \sqrt{m_0 \left(\varepsilon+m_0\right)+v^2 k_y (k_y-\kappa)}
\nonumber \\ \times
\left(\varepsilon +m_0-iv(\kappa + k_y) \frac{k_+}{k}\right) \hskip1cm
\nonumber \\ \times
\left[ \left(\varepsilon +m_0-iv(\kappa + k_y) \frac{k}{k_-}\right) \hskip1cm
\right. \nonumber \\ \left. \times
\left(\varepsilon +m_0 - i v (\kappa + k_y) \frac{k_+}{k}\right) \hskip1cm
\right. \nonumber \\ \left.
-e^{2\kappa L}\, \left(\varepsilon +m_0 + iv(\kappa + k_y)\frac{k}{k_-}\right) \hskip1cm
\right. \nonumber \\ \left. \times
\left(\varepsilon +m_0 + i v (\kappa - k_y) \frac{k_+}{k}\right) \right]^{-1}.
\end{eqnarray}
The results (15) and (17) can be used for the calculation of various transport properties of
TI/magnetic layer structures. Here we present the calculation of conductance and thermoelectricity.

\section{\label{sec:level1}Conductance}

\begin{figure}
	\vspace*{-0.8cm}
	\hspace*{-1cm}
	\includegraphics[width=9cm]{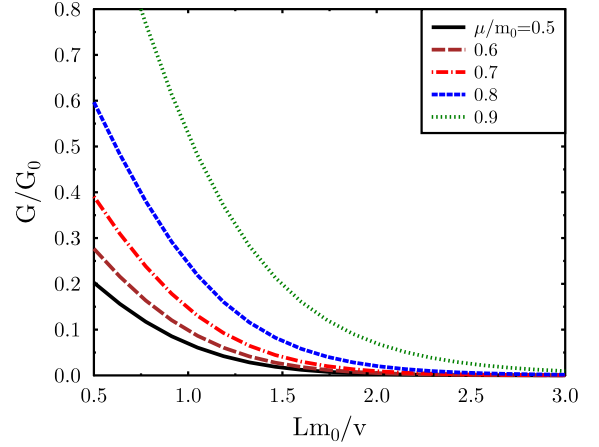}
	\hspace*{-1cm}
	\includegraphics[width=9cm]{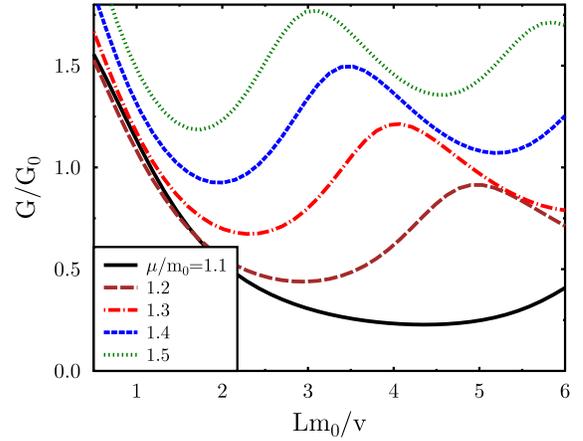}
	\vspace*{-0.2cm}
	\caption{Electrical conductance as a function of length $L$ of the magnetized region for indicated values of the Fermi level $\mu $.
		The character of curves is different in the tunneling regime, $\mu <m_0$ and at the overbarrier transmission, $\mu >m_0$.}
\end{figure}

Now we calculate the conductance of the structure assuming that the magnetic nanoribbon is a unique
barrier for the motion of electrons along the surface (usual impurities are fully transparent).
Then we take chemical potential $\mu $ on the right side (region III) and $\mu +eU$ on the left side
(region I). This corresponds to the electron flux from left to right (see Fig.~1) under
external voltage $U$.

The electric current across the magnetic region reads
\begin{eqnarray}
\label{18}
j=\frac{ev}{\hbar }{\sum }'_{\bf k}\frac{k_x}{k} |t({\bf k})|^2
\big[ \theta (\mu +eU-\varepsilon _{1k})
-\theta (\mu -\varepsilon _{1k})\big] ,\hskip0.4cm
\end{eqnarray}
where $vk_x/\hbar k$ is the $x$-component of electron velocity in region III,
$\varepsilon _{1k}=vk$ is the energy spectrum of electrons in region III
with energy $\varepsilon >0$, and $\theta (z)$ is
the Heaviside's function. The sum over ${\bf k}$ is restricted by $k_x>0$.

The results of numerical calculation of the electric conductance $G=jL/U$ are presented in Fig.~4, where
we denote $G_0=e^2/\hbar $. As we see, for $\mu >m_0$ the conductance is also oscillating with $L$, similarly to the
transmission probability $T_{k_y=0}$. This is because the main contribution to the current is related
to electrons incoming with the small value of $k_y$, i.e., perpendicular to the interface.

\section{\label{sec:level1}Thermoelectricity}

\begin{figure}
	\vspace*{-0.7cm}
	\hspace*{-1cm}
	\includegraphics[width=9cm]{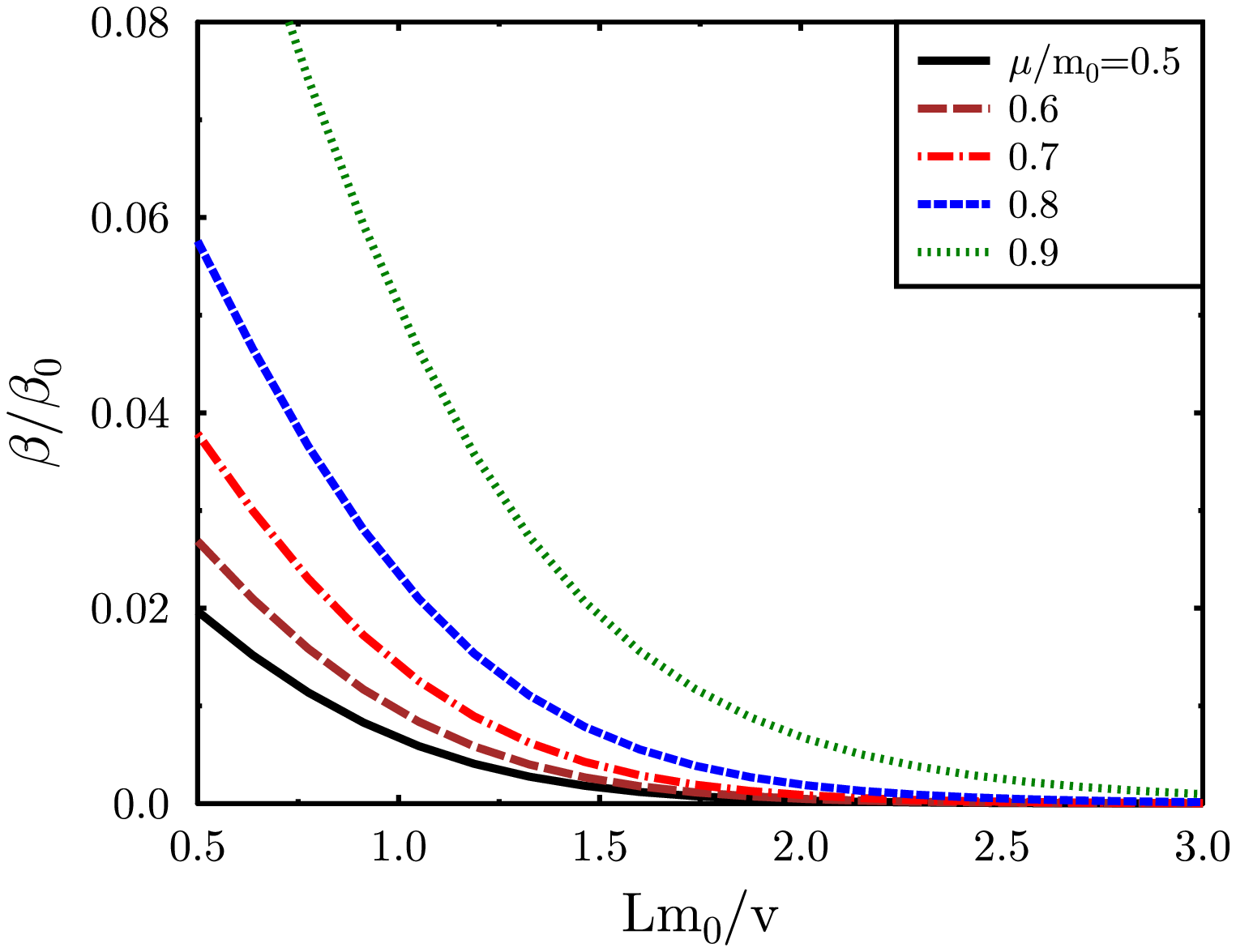}
	\hspace*{-1cm}
	\includegraphics[width=9cm]{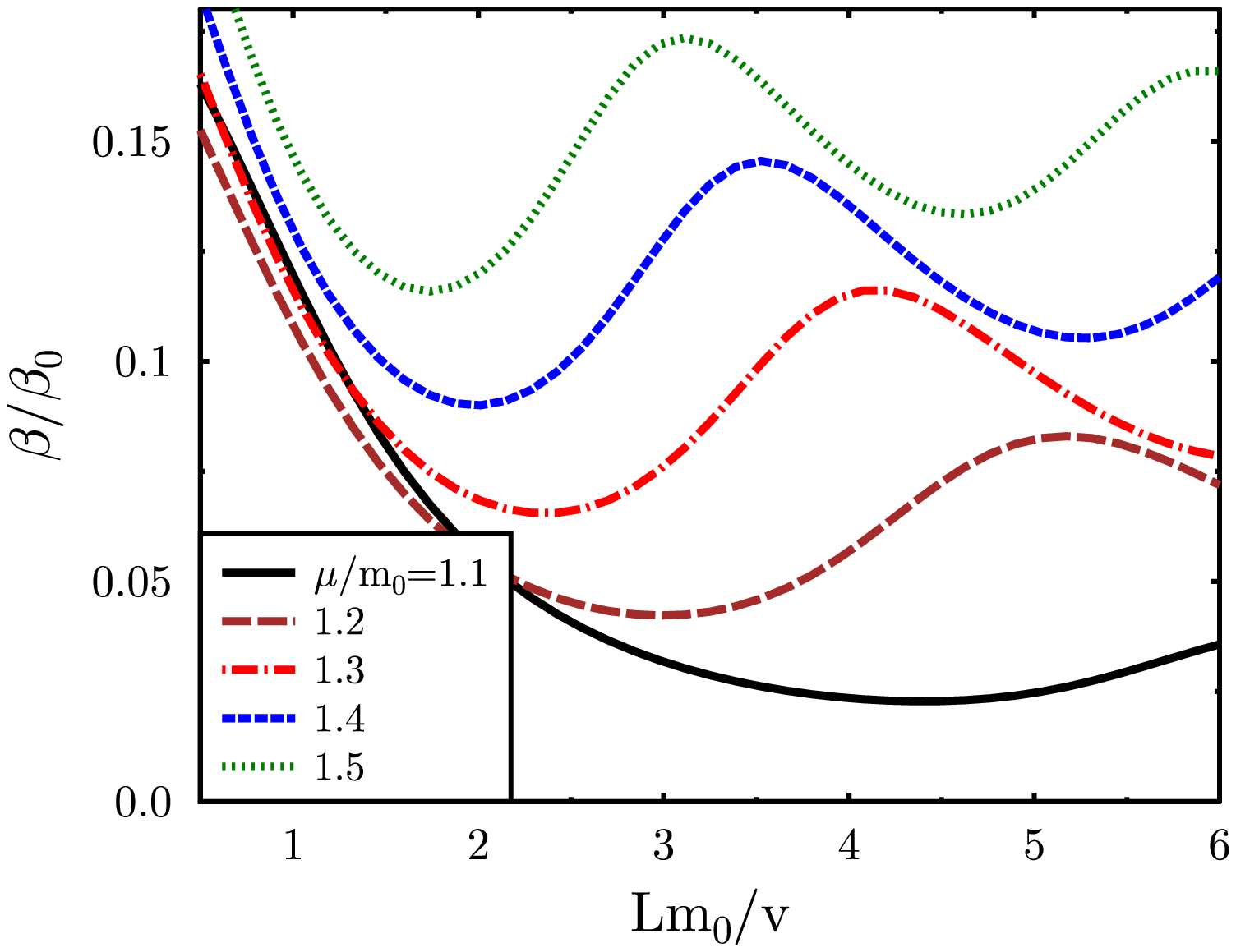}
	\vspace*{-0.2cm}
	\caption{Linear response thermoelectric coefficient $\beta $ as a function of $L$ for different values of the chemical potential $\mu $.}
\end{figure}

Using the transmission coefficient (15) and (17) in the region of overbarrier motion and tunneling, 
one can also calculate the electric current generated by
a temperature gradient (Seebeck effect). We assume that the temperature of  TI in the regions I and III
is $T_{1,2}=T\pm \Delta T/2$, respectively. The thermoelectric current can be  then written in the form
\begin{eqnarray}
\label{19}
j_T=\frac{ev}{\hbar }{\sum }'_{\bf k}\frac{k_x}{k}\, |t({\bf k})|^2
\big[ f_1(\varepsilon _{1k}-\mu )-f_2(\varepsilon _{1k}-\mu )\big] , \hskip0.4cm
\end{eqnarray}
where $f_{1,2}(z)=[\exp (z/k_BT_{1,2})+1]^{-1}$ are the  Fermi-Dirac distribution functions in the region I and III,
and we assume $\Delta T\ll \mu /k_B$ so that the thermoelectricity is due to electrons in the upper
energy band $\varepsilon _{1k}$.

The results of numerical calculation of the linear response thermoelectric coefficient $\beta$ defined as
$\beta \equiv GS=j_TL/m_0T$,
where $S$ is the Seebeck coefficient and we denote $\beta _0=ek_B/\hbar $,  are presented in Fig.~5.
Like the electrical conductance, the thermoelectric coefficient $\beta $ reveals some oscillations
with increasing magnetic region length $L$. As we see, the dependence of $\beta $ on $L$ is very similar to
the dependence of conductance $G(L)$ because in both cases the main effect is mostly related to the dependence of
transmission coefficion on the length $L$.

\section{\label{sec:level1}Application of the Dirac model to ${\rm Bi}_2{\rm Se}_3$ topological insulator and the spin torque}

The standard model of ${\rm Bi}_2{\rm Se}_3$ topological insulator uses the following Hamiltonian (without
magnetization term)
\begin{eqnarray}
\label{20}
H_0=-iv(\sigma _x\partial _y-\sigma _y\partial _x),
\end{eqnarray}
which differs from the Dirac model in Eq.~(1). However, by using unitary transformation $U={\rm diag }(1,i)$
the Dirac Hamiltonian $H_0=-iv \mbox{$\boldsymbol{\sigma}$} \cdot \mbox{$\boldsymbol{\nabla}$} $ transforms into (20). It means just a different choice of the phase in basis functions
in "different" models. The corresponding transformation of Pauli matrices is $U^\dag \sigma _xU=-\sigma _y$,
$U^\dag \sigma _yU=\sigma _x$ and $U^\dag \sigma _zU=\sigma _z$. Thus, the Dirac model with magnetization (1)
describes the Bi$_2$Te$_3$ TI with magnetization along the same axis $z$. The spin
polarization of electrons along the direction of motion (helicity) in the Dirac model corresponds to the polarization perpendicular
to the direction of electron motion in Bi$_2$Te$_3$ TI. This allows us to use all the results of calculation
of transmission of electrons through the magnetized region in Bi$_2$Te$_3$ with the standard Dirac model (1).

The problem of spin torque needs special consideration, which is not in the scope of this paper,
and we restrict ourselves by briefly discussing the main points. The total spin transfer torque (STT)
acting on the moments in the magnetized regions can be calculated
as a difference of incoming and outgoing spin currents. Since the spin currents in incoming
and reflected waves have the same sign, and there is a relation between reflection $R$ and transmission $T$
probability, $T+R=1$, one can find the following expression for STT
\begin{eqnarray}
\label{21}
\mathcal{T}_{STT}=v {\sum }'_{\bf k}\frac{k_x}{k}\, \big( 1-|t({\bf k})|^2\big)
\big[ \theta (\mu +eU-\varepsilon _{1k})
\nonumber \\
-\theta (\mu -\varepsilon _{1k})\big] .
\end{eqnarray}
In the case of Bi$_2$Te$_3$ TI the spin polarization of current is along $-y$.
Thus, the STT component of torque is also along $-y$ leading to current-induced deviation
of magnetic moments from perpendicular.
The maximum of STT is for $|t({\bf k})|^2=0$ (full reflection), whereas at a full transparency,
$|t({\bf k})|^2=1$, STT is zero.

Another component of the torque, namely, the spin-orbit torque (SOT) is nonzero due to the
spin-polarized electrons in TI.
Indeed, the nonequilibrium current along axis $x$ is associated with
the prevailing number of electrons with spin in $-y$ direction, which results in a nonequilibrium
magnetization $\mathcal{M}_y$ of the electron gas. In its turn, it leads to the generation of SOT acting
at the magnetic moments as $ \mbox {$\boldsymbol{\mathcal{T}}$} _{SOT}\sim {\bf M}\times \mbox {$\boldsymbol{\mathcal{M}}$} $, i.e. along $x$.

\section{Conclusions}

We calculated the conductance and thermoelectric coefficient of a topological-insulator-based structure
consisting of a TI and  a thin magnetic ribbon on its surface. In fact, these parameters characterize transport
properties of the region II since transport in the regions I and III is purely ballistic.

It should be noted that we did not take into account the possibility of scattering from impurities
in the region II. In contrast to the transport through regions I and III, transport in the region II
depends on the presence of potential scatterers, which appears due to the gap. However, we can neglect this
effect as we consider narrow magnetic ribbon, i.e., when $L\ll \ell $, where $\ell $ is the mean free path of electrons
in the region $0<x<L$.

It is also interesting to analyze the problem of spin transmission through the region under the magnetic ribbon
in the Dirac model of TI.
Generally, the electron spin is not conserved when an electron
traverses the magnetic region, contrary to the electric charge which has to be conserved.
As we see from Eqs.~(4) and (5), spin of an electron changes its orientation when it
passes through  the interface $x=0$, whereas reflected electrons have spin orientation opposite to the spin of
incoming electrons.
Thus, the flux of electrons reflected from the interface $x=0$ is associated with
the spin current $J_x$ (spin along $x$) in the same direction as the spin flux of incoming electrons.
This indicates that the spin current in region I
is enhanced with increasing reflection coefficient $R=|r|^2$. In the limit of $R\to 1$ we get the net spin current
generator in the absence of charge current.
In turn, the spin current in the region III flows in the same direction as the particle (electron) flux, which is proportional
to the transmission coefficient $T=|t|^2$. Therefore, the spin current passing through the structure is not conserved,
due to the spin torque exerted on magnetic moments in the region II. The maximum spin torque corresponds
to the limit of $T=0$, i.e., to the limit of total reflection of electrons.

\begin{acknowledgments}
	This work was supported by the National Science Center in Poland under the Project No.~DEC-2012/04/A/ST3/00372.
SK aknowledges support of the National Science Center in Poland under Project No.~DEC-2012/04/A/ST3/00042.
The authors thank J. Barna\'s for critical reading of the manuscript and useful comments.
\end{acknowledgments}


\begin{thebibliography}{[1]}
	
	\bibitem{hasan10}
	M. Z. Hasan and C. L. Kane,
	Rev. Mod. Phys. {\bf 82}, 3045 (2010),
	
	\bibitem{qi11}
	X. L. Qi and S. C. Zhang,
	Rev. Mod. Phys. {\bf 83}, 1057 (2011).
	
	\bibitem{fu06}
	L. Fu and C. L. Kane,
	Phys. Rev. B {\bf 74}, 195312 (2006).
	
	\bibitem{fu07}
	L. Fu and C. L. Kane,
	Phys. Rev. B {\bf  76}, 045302 (2007),
	
	\bibitem{liu09}
	Q. Liu, C. X. Liu, C. Xu, X. L. Qi, and S. C. Zhang,
	Phys. Rev. Lett. {\bf 102}, 156603 (2009).
	
	\bibitem{zitko10}
	R. ${\rm \check{Z}}$itko,
	Phys. Rev. B {\bf 81}, 241414 (2010).
	
	\bibitem{abanin11}
	D. A. Abanin and D. A. Pesin,
	Phys. Rev. Lett. {\bf 106}, 136802 (2011).
	
	
	\bibitem{cheianov12}
	V. Cheianov, M. Szyniszewski, E. Burovski, Yu. Sherkunov, and V. Fal'ko,
	Phys. Rev. B {\bf 86}, 054424 (2012).
	
	\bibitem{rosenberg12}
	G. Rosenberg and M. Franz,
	Phys. Rev. B {\bf 85}, 195119 (2012).
	
	\bibitem{biswas10}
	R. R. Biswas and A. V. Balatsky,
	Phys. Rev. B {\bf 81}, 233405 (2010).
	
	\bibitem{zhang12}
	D. Zhang, A. Richardella, D. W. Rench, S. Y. Xu, A. Kandala, T. C. Flanagan, H. Beidenkopf, A. L. Yeats,
	B. B. Buckley, P. V. Klimov, D. D. Awschalom, A. Yazdani, P. Schiffer, M. Z. Hasan, and N. Samarth,
	Phys. Rev. B {\bf 86}, 205127 (2012).
	
	\bibitem{henk12}
	J. Henk, A. Ernst, S. V. Eremeev, E. V. Chulkov, I. V. Maznichenko, and I. Mertig,
	Phys. Rev. Lett. {\bf 108}, 206801 (2012).
	
	\bibitem{chen10}
	Y. L. Chen, J.-H. Chu, J. G. Analytis, Z. K. Liu, K. Igarashi, H.-H. Kuo, X. L. Qi, S. K. Mo, R. G. Moore,
	D. H. Lu, M. Hashimoto, T. Sasagawa, S. C. Zhang, I. R. Fisher, Z. Hussain, Z. X. Shen,
	Science {\bf 329}, 659 (2010).
	
	\bibitem{wray11}
	L. A. Wray,	S.-Y. Xu, Y. Xia, D. Hsieh,	A. V. Fedorov,	Y. S. Hor, R. J. Cava, A. Bansil, H. Lin and M. Z. Hasan,
	Nature Phys. {\bf 7}, 32 (2011).
	
	\bibitem{katsnelson}
	M. I. Katsnelson, Graphene: Carbon in Two Dimensions, Chap.~4 (Cambridge Univ. Press, 2012).
	
	\bibitem{kong11}
	B. D. Kong, Y. G. Semenov, C. M. Krowne, and K. W. Kim,
	Appl. Phys. Lett. {\bf 98}, 243112 (2011).
	
	\bibitem{yokoyama11}
	T. Yokoyama,
	Phys. Rev. B {\bf 84}, 113407 (2011).
	
	\bibitem{fan14}
	Y. Fan, P. Upadhyaya, X. Kou, M. Lang, S. Takei, Z. Wang, J. Tang, L. He, L.-T. Chang, M. Montazeri,
	G. Yu, W. Jiang, T. Nie, R. N. Schwartz, Y. Tserkovniak, and K. L. Wang,
	Nature Mat. {\bf 13}, 699 (2014).
	
	\bibitem{mellnik14}
	A. R. Mellnik, J. S. Lee, A. Richardella, J. L. Grab, P. J. Mintun, M. H. Fischer, A. Vaezi, A. Manchon, E.-A. Kim,
	N. Samarth, and D. C. Ralph,
	Nature {\bf 511}, 449 (2014).
	
	\bibitem{semenov14}
	Y. G. Semenov, X. Duan, and K. W. Kim,
	Phys. Rev. B {\bf 89}, 201405(R) (2014).
	
	\bibitem{chang15}
	P.-H. Chang, T. Markussen, S. Smidstrup, K. Stokbro, and B. K. Nikoli\'c,
	Phys. Rev. B {\bf 92}, 201406(R) (2015).
	
	\bibitem{mahfouzi16}
	F. Mahfouzi, B. K. Nikoli\'c, and N. Kioussis,
	Phys. Rev. B {\bf 93}, 115419 (2016).
	
	\bibitem{kurebayashi14}
	H. Kurebayashi, J. Sinova, D. Fang, A. C. Irvine, T. D. Skinner, J. Wunderlich, V. Nova\'ak, R. P. Gallagher, E. K. Vehstedt,
	L. P. Z\'arbo, K. V\'yborn\'y, A. J. Ferguson, and T. Jungwirth,
	Nature Nanotechnol. {\bf 9}, 211 (2014).
	
	\bibitem{tanaka09}
	Y. Tanaka, T. Yokoyama, and N. Nagaosa, 
	Phys. Rev. Lett. {\bf 103}, 107002 (2009).
	
	\bibitem{dolcini15}
	F. Dolcini, M. Houzet, and J. S. Meyer,
	Phys. Rec. B {\bf 92}, 035428 (2015).
	
	
	
	
	
	
	
	
	
	
	
	
	
	
	
	
	
	
	
	
	
	
	
	
	
	
	
	
\end{thebibliography}
\end{document}